# Search for Intra-Cluster Machos by Pixel Lensing of M87

Andrew Gould[1]

Dept of Astronomy, Ohio State University, Columbus, OH 43210

e-mail gould@payne.mps.ohio-state.edu

## Abstract

Intra-Cluster Machos (ICMs) are a plausible candidate for at least some of the dark matter in clusters of galaxies. ICMs can be detected by searching toward M87 for "pixel lensing", gravitational microlensing of unresolved stars. Dedicated observations by the Wide Field and Planetary Camera on the *Hubble Space Telescope* would discover lenses at a rate $\Gamma \sim 7f\,\mathrm{day}^{-1}$ where $f$ is the fraction of the Virgo cluster composed of Machos (Massive Compact Objects). The only important background is lensing by stars in M87 itself. Such self-lensing should be detected at a rate $\Gamma \sim 1\,\mathrm{day}^{-1}$. However, most of these background events should be recognizable from their angular distribution and/or measurement of their proper motions (which are smaller by a factor $\sim 2$ than those of ICMs). The observations should be carried out principally in $I$, but with $V$ measurements made $\sim 20\%$ of the time in order to determine the colors of the lensed stars. It is also possible to detect "Class II" lensing events which are longer and hence for which lower signal to noise is balanced by a much larger number of observations per event. There are $\sim 5$ times more of these than the higher quality events. However, it is difficult to distinguish Class II events generated by ICMs from those generated by M87 stars.

Subject Headings: galaxies: individual: M87 – galaxies: clusters: individual: Virgo – gravitational lensing

## 1. Introduction

Large quantities of previously dark baryonic matter has been found in clusters of galaxies in the form of hot X-ray gas. In particular White (1993) has used X-ray observations to measure both the hot baryonic component and the total mass of the core of the Coma cluster. The ratio of these quantities gives a lower limit to the baryonic mass fraction of Coma and so, if Coma is representative, of the universe as a whole. When this ratio is combined with nucleosynthesis arguments constraining $\Omega_b$, it then places an upper limit on $\Omega$, the density of the universe as a fraction of closure density. In order to transform these limits into measurements, one would have to determine how much dark baryonic matter is in clusters apart from the X-ray gas.

Machos (Massive Compact Objects) are the most plausible baryonic candidate which might contribute significant additional mass to clusters. Two ongoing experiments are searching for Machos in the Galactic halo by monitoring $\sim 10^7$ stars in the Large Magellanic Cloud (LMC). If Machos made up a standard dark halo then at any given time a fraction $\tau \sim 5 \times 10^{-7}$ of these stars would be significantly magnified by a foreground Macho along the line of sight. The two groups obtain fairly similar results consistent with an optical depth $\tau \sim 10^{-7}$ (Alcock et al. 1993, 1995b MACHO; Aubourg et al. 1993 EROS). MACHO has shown that their result rules out a standard dark halo of Machos. However, the observed optical depth is much larger than would be expected from known stellar components in the disk and thick disk (Gould, Bahcall, & Flynn 1995), the spheroid (Bahcall et al. 1994) and the LMC (Gould 1995a). Thus, the events may well be due to an extended halo of Machos which accounts for only a fraction of the total dark matter. If so, Alcock et al. (1995b) have shown that the total mass in Machos inside the orbit of the LMC is $\sim 7 \times 10^{10} \, M_\odot$ and is fairly independent of the parameters adopted for the halo model. This mass is of order the total baryonic material in the Milky Way disk.

One might then speculate as follows about the history of baryonic matter in clusters. At the stage where proto-galaxies are forming, those in the outskirts of the cluster begin evolving similarly to the Milky Way: half of their gas is processed into a partial baryonic halo of Machos while the remaining gas forms a proto-disk. Before this disk can form a significant number of stars, the galaxy falls deep into the cluster potential well and the gas is stripped to contribute to the cluster X-ray halo. The Machos continue to orbit in the cluster, either bound to one another in a relatively distinct dark galaxy or possibly dispersed by interactions with other galaxies. In either case, one would expect the cluster to contain approximately equal masses in X-ray gas and Machos.

Here I propose a method to search for such Intra-Cluster Machos (ICMs) by pixel lensing, a method now being pioneered by two groups (Crotts 1992; Baillon et al. 1993) in their searches for Machos toward M31. I show that by continuously monitoring M87 with the Wide Field and Planeterary Camera (WFPC2) on the *Hubble Space Telescope (HST)*, one could detect ICMs in Virgo at a rate

$$\Gamma \sim 7f \, \text{day}^{-1} \qquad (1.1)$$

where $f$ is the fraction of intra-cluster dark matter in Virgo in the form of ICMs. Thus, for $f \sim 20\%$ (similar to the baryonic gas fraction in Coma) one could expect to detect about 1.5 events per day.

In § 2, I develop a general formalism which enables one to understand all classes of pixel lensing experiments. In § 3, I apply this formalism to *HST* observations of M87 to derive equation (1.1). In § 4, I discuss various foregrounds and backgrounds and show that the principal background is lensing by stars in M87 itself. In § 5, I discuss the measurement of Macho proper motions and show how these can be used to distinguish between the intra-cluster signal and the M87 background. In § 6, I discuss "Class II" lensing events which are $\sim 5$ times more numerous than primary events discussed in the paper. Finally, I summarize my conclusions in § 6.

## 2. Pixel Lensing

In standard lensing searches of the type proposed by Paczyński (1986) and currently being carried out by MACHO and EROS toward the LMC and by MACHO (Alcock et al. 1995a) and OGLE (Udalski et al. 1994) toward the Galactic bulge, one monitors individual stars for time-dependent magnifications

$$A[x(t)] = \frac{x^2 + 2}{x(x^2 + 4)^{1/2}}, \tag{2.1}$$

where $x$ is the projected separation between the source and the lens normalized to the Einstein radius

$$\theta_* = \left(\frac{4GMD_{\rm LS}}{c^2 D_{\rm OL} D_{\rm OS}}\right)^{1/2} \tag{2.2}$$

and where $M$ is the Macho mass and $D_{\rm OL}$, $D_{\rm OS}$, and $D_{\rm LS}$ are the distances between the observer, source, and lens. For rectilinear motion, the projected separation is given by

$$x(t) = \sqrt{\omega^2(t - t_0)^2 + \beta^2}, \tag{2.3}$$

where $t_0$ is the time of maximum magnification, $\beta$ is the impact parameter normalized to $\theta_*$ and $\omega^{-1}$ is the characteristic time of the event,

$$\omega^{-1} = \frac{\theta_* D_{\rm OL}}{v}. \tag{2.4}$$

Here $v$ is the transverse speed of the Macho relative to the observer-source line of sight.

However, it is not in general possible to monitor individual stars in more distant galaxies such as M31 because few stars are resolved. Instead, Crotts (1992) and Baillon et al. (1993) have begun searching for lensing events of unresolved stars in M31 by measuring the time dependence of the flux in individual pixels (or groups of pixels) against the general background light. I refer to such time-dependent fluxes as "pixel lensing". A major technical concern in these searches is how to account for seeing variation. The two groups are approaching this problem differently and the ultimate viability of the experiments will depend on how well they are able to deal with it. In the present work, however, I am proposing a space-based pixel lensing search of M87. Since seeing variation is negligible from space, I will ignore this problem.

Suppose that one is monitoring a galaxy with surface brightness $S$ using a CCD detector and with angular area $\Omega_{\rm ccd}$ with a point spread function (PSF) of characteristic area $\Omega_{\rm psf}$. [That is, $\Omega_{\rm psf}$ is defined so that the background noise for faint point-source photometry is given by $\Omega_{\rm psf} B$, where $B$ is the total number of background counts per unit area. For example, for a well-sampled Gaussian PSF with FWHM $\theta_{see}$, $\Omega_{\rm psf} = (\pi/\ln 4)\theta_{see}^2$.] And suppose that a lensing event is recognizable provided that the peak magnification produces a change in the luminosity of an unresolved star of at least $\Delta L_{\min}$. Then, as I will show below, the lensing event rate is given by

$$\Gamma = \Gamma_0 \frac{\Omega_{\rm ccd}}{\Omega_{\rm psf}} \eta \xi, \qquad \Gamma_0 \equiv \frac{2}{\pi} \langle \omega \tau \rangle \qquad (2.5)$$

where $\xi$ is the fraction of the galaxy's luminosity due to 'lensable stars' (defined more precisely below), and

$$\eta \equiv \frac{S \Omega_{\rm psf}}{\Delta L_{\min}}. \qquad (2.6)$$

Note that $\Gamma_0$ is just the lensing rate for a single resolved star.

For simplicity, I initially assume that all stars in the galaxy are point sources, that the interval between observations is short compared to other time scales in the problem, and that stars with luminosity $L \gtrsim \Delta L_{\min}$ do not contribute significantly to the integrated luminosity of the galaxy. Later I will relax these assumptions. However, I note that all three assumptions do approximately apply to the proposed observations of M87. Consider then a Macho with Einstein radius $\theta_*$ and a source star with luminosity $L_i \ll \Delta L_{\min}$. The lensing event will be observable only if the peak magnification obeys $A_{\max} > \Delta L_{\min}/L_i$, where $A_{\max}$ is the maximum magnification during the event. From equation (2.1) this will occur provided $\beta < L_i/\Delta L_{\min}$. Hence, the (angular) cross section for lensing of this star is $2\beta\theta_* = 2(L_i/\Delta L_{\min})\theta_*$. The total cross section over all stars in the field is then

$$\theta_{\text{tot}} = \sum_i 2\frac{L_i}{\Delta L_{\min}}\theta_* = 2\frac{S\Omega_{\text{ccd}}}{\Delta L_{\min}}\theta_*. \tag{2.7}$$

Consider now an angular density of lenses, $n$, all with the same $\theta_*$ and proper motion $\mu = \omega\theta_*$. The lensing rate is then given by

$$\Gamma = n\theta_{\text{tot}}\mu = \frac{2}{\pi}\omega\tau\frac{S\Omega_{\text{ccd}}}{\Delta L_{\min}}, \tag{2.8}$$

where $\tau = \pi\theta_*^2 n$ is the optical depth. If one now considers an ensemble of lenses with different $\theta_*$ and $\mu$, equation (2.8) remains valid provided that $\omega\tau \to \langle\omega\tau\rangle$. Notice that equation (2.8) is identical to equation (2.5) with $\xi = 1$. That is, equation (2.5) is valid for the case when the entire luminosity function contributes to lensing.

There are two classes of stars which might not contribute to lensing according to their total luminosity: low luminosity stars which cannot be magnified enough to be seen and high luminosity stars $L \gtrsim \Delta L_{\min}$ for which the cross section is simply $2\theta_*$ rather than the luminosity-weighted $2(L/\Delta L_{\min})\theta_*$ which enters equation (2.7). The low luminosity stars are eliminated whenever the required magnification $\Delta L_{\min}/L$ is so large and the corresponding impact parameter $\beta$ is so small, that the Macho transits the face of the star, i.e., $L \lesssim \Delta L_{\min}(\theta_s/\theta_*)$ where $\theta_s$ is the angular radius of the star. In this case, the star will be magnified by less than the amount predicted by the point-source formula (2.1) so that it will not in fact be seen. Also, if the sampling intervals are too long, the peak might be missed so that the event would also not be seen. I denote this minimum stellar luminosity $L_{\min}$. On the other hand, if stars are so bright that their magnification can be measured even when $A \sim 1$, then equation (2.7) will drastically overestimate their contribution.

They should therefore be ignored in the summation over the luminosity function. Hence, the true lensing rate should be reduced by a factor

$$\xi = \int_{L_{\min}}^{\Delta L_{\min}} dL\phi(L)L \Big/ \int_0^\infty dL\phi(L)L, \qquad (2.9)$$

where $\phi(L)$ is the luminosity function. This completes the proof of equation (2.5).

For a fixed mass $M$ and fixed mean transverse speed $\bar{v}$, $\Gamma_0$ can be evaluated

$$\Gamma_0 = \frac{2}{\pi}\langle\omega\tau\rangle = 2\bar{v}\int_0^{D_{\rm OS}} dD_{\rm LS}\rho(D_{\rm LS})\sqrt{\frac{4GD_{\rm OL}D_{\rm LS}}{Mc^2 D_{\rm OS}}}. \qquad (2.10)$$

For a line of sight to the center of an isothermal sphere characterized by rotation speed $v_c$ and core radius $a$, and assuming $D_{\rm LS} \ll D_{\rm OS}$, I find

$$\Gamma_0 = \left(\frac{\pi}{2}\right)^{1/2}\frac{v_c^3}{c^2}\left(\frac{4GMa}{c^2}\right)^{-1/2}. \qquad (2.11)$$

## 3. Application to M87

I now apply equation (2.5) to observations of M87. For definiteness, I adopt Virgo-cluster parameters $v_c \sim 1400\,{\rm km\,s^{-1}}$ (corresponding to a one-dimensional dispersion of $\sim 1000\,{\rm km\,s^{-1}}$), core radius $a \sim 250\,{\rm kpc}$, and $M = 0.1\,M_\odot$. I then find from equation (2.11) that,

$$\Gamma_0 = 1.6 \times 10^{-6} f\,{\rm day}^{-1}, \qquad (3.1)$$

for a Virgo cluster with Macho fraction $f$. I use the adopted parameters to estimate typical time scales and Einstein radii,

$$\omega^{-1} \sim 16\,{\rm day}; \qquad \theta_* D_{\rm OS} \sim 14\,{\rm AU}. \qquad (3.2)$$

I also adopt a distance modulus to Virgo of 31. As I will discuss below, the optimal bandbass is $I$. J. Tonry (private communication 1995) has measured the surface brightness of M87 in $I$ as a function of angular radius and finds values of 16, 17, 18, 19, 20, and 21 mag arcsec$^{-2}$ at $3''$, $12''$, $24''$, $41''$, $71''$, and $114''$ respectively. This means that the *HST* WFPC2 field (three $800 \times 800$ chips with $0\overset{''}{.}1$ pixels and one chip with $0\overset{''}{.}044$ pixels) could image a region with surface brightness in the range $16 - 21$ mag arcsec$^{-2}$ and with a typical value of 19.5 mag arcsec$^{-2}$.

For the three chips with large pixels, the effective area of the PSF will depend on where the lensed star falls on the pixel. Based on past experience (Bahcall et al. 1994) I adopt $\Omega_{\rm psf} = 0.04\,\rm arcsec^2$, i.e., 4 WFC2 pixels. The chip with smaller pixels (PC2) has a somewhat smaller $\Omega_{\rm psf}$ but I adopt the same value to be conservative and in the interests of simplicity. Assuming that $760^2$ pixels of each chip are usable, I find $\Omega_{\rm ccd} = 5.1\,\rm arcmin^2$. Hence,

$$\frac{\Omega_{\rm ccd}}{\Omega_{\rm psf}} = 4.6 \times 10^5. \tag{3.3}$$

I now assume that observations are carried out twice per orbit ($\sim 90$ min) each for 900 s in I band (F814W). I assume that to be detectable, the signal to noise at peak must be $\gtrsim 6$ in the image formed by combining these two exposures. I then analyze the observability of lensing events at three values of $S$, the typical 19.5 mag arcsec$^{-2}$ and 1.5 mag above and below this value. At the typical value, 3500 photons would fall on $\Omega_{\rm psf}$ during the 1800 s of exposures during one orbit. Hence, a 6 $\sigma$ detection would require a 10% fluctuation or a $\Delta L_{\min}$ corresponding to $I \sim 25.5$. Stars near the base of the giant branch $M_I \sim 0.5$ have $I \sim 31.5$ and so must be magnified 160 times to be detected. This corresponds to an impact parameter $\sim 19 R_\odot$ which is considerably larger than the size of such stars $\sim 9\,R_\odot$. Hence little of the total luminosity lies in stars that are too faint to be effectively lensed. Similarly, one finds that the characteristic time of the event $\beta/\omega \sim 2.4\,\rm hr$ is substantially longer than the sampling time, so the event could easily be resolved in time. On the other hand, in order for a star to be so bright that it could be noticed even for low magnifications $A \sim 1$, it would have to have $M_I \lesssim -5.5$. The fraction of the total luminosity due to such stars is negligibly small. Hence at $S = 19.5 \, \mathrm{mag \, arcsec^{-2}}$, I find $\eta \sim 10$ and $\xi \sim 1$. Repeating the calculation at 1.5 mag (i.e., a factor $\sim 4$) fainter, I find that $\eta$ is reduced by a factor $\sqrt{4}$ to $\eta \sim 5$. At the faint end of the luminosity function the required magnification is a factor $\sim 2$ smaller, so that there is even less problem with the finite size of the stars. The bright cutoff is now $M_I \sim -4.75$, still too bright to have a major effect. Hence $\xi \sim 1$. Finally, repeating the calculation at $S = 18 \, \mathrm{mag \, arcsec^{-2}}$, I find that $\eta \sim 20$. At the faint end of the luminosity function, required impact parameters are $\sim 10 \, R_\odot$ implying that they marginally satisfy the constraint of being larger than the star. I adopt $\xi \sim 0.8$. Over the entire field I therefore estimate $\eta \xi \sim 10$.

Combining these results with equations (2.5), (3.1), and (3.3), I estimate an event rate for the Virgo cluster of $\Gamma \sim 7f \, \mathrm{day}^{-1}$.

In making the above calculations, I have focused attention on the base of the giant branch. In doing so, I implicitly assumed that if these relatively faint stars could give rise to observable lensing, then brighter stars could also. In effect, I have assumed that as one moves up the giant branch, luminosity ($L$) grows at least as fast as radius ($R$). Recall that the critical impact parameter is given by $\beta = L/\Delta L_{\min}$. If this grew more slowly than the radius then lenses which were just able to generate an observable event on a faint star would be unable to do so on a bright star. For the $I$ band the assumption that $L/R$ rises in fact holds. For example I find for a 10 Gyr metal-rich population, that when $L$ rises by a factor 7.5, $R$ rises by a factor $\sim 4.1$ (Green, Demarque, & King 1987). Hence lensing becomes slightly more feasible for brighter stars. However, in $V$ band, $L/R$ tends to decline for brighter giants. This, taken together with the fact that metal-rich giants are a red population and hence easier to observe in $I$ than $V$ makes $I$ the optimal optical band.

## 4. Foregrounds and Backgrounds

There are several possible instrumental, stellar, and lensing backgrounds which could in principle mimic the expected signal from the ICMs. By far the most important contaminant is lensing by stars in M87 itself. Adopting an isothermal-sphere model with one-dimensional dispersion $\sim 360\,\mathrm{km\,s^{-1}}$ (Faber et al. 1989), taking the typical stellar mass to be $M \sim 0.2\,M_\odot$, and applying the results of Gould (1995c), I find

$$\Gamma_0 = 5 \times 10^{-7} (M/0.2 M_\odot)^{-1/2} (\theta/100'')^{1/2} \,\mathrm{day}^{-1} \qquad (\text{M87 stars}). \qquad (4.1)$$

This is comparable to the rate from the ICMs for $f \sim 30\%$. Note, however, that the typical Einstein ring of the M87 stars at $\theta = 100''\,(\sim 8000\,\mathrm{pc})$ is a factor $\sim 3$ smaller than the typical Einstein ring of the ICMs. This implies that the M87 stars are much more susceptible to finite source size effects than are the ICMs. Reviewing the discussion in the previous section, I find that the maximum impact parameters for source stars with $M_I = 0.5$ to be lensed by other M87 stars are $13\,R_\odot$, $5\,R_\odot$, and $1.6\,R_\odot$ at $114''$, $56''$, and $24''$ respectively. Since the radii of such sources are $\sim 9\,R_\odot$, the effects in the three regimes are modest, significant, and severe respectively. In the inner regions, therefore, the M87 stellar events would either go unnoticed or if they were seen, would often be recognized as having exceptionally small Einstein rings because of stellar transits (see § 5). Thus, the ICMs would be relatively free from contamination by M87 stars in the inner part of the galaxy, but probably not in the outer part.

A second and much less severe source of contamination comes from foreground lensing by Machos in the Galactic halo. Both the MACHO and EROS experiments find an optical depth $\tau \sim 10^{-7}$ and $\langle \omega \rangle \sim (20\,\mathrm{day})^{-1}$ for observations toward the LMC. This leads to an estimate $\Gamma_0 \sim 3 \times 10^{-9}\,\mathrm{day}^{-1}$, a factor $(500f)^{-1}$ lower than the Virgo rate. Unless $f$ is unmeasurably small, the Galactic rate is completely negligible.

There are two possible instrumental backgrounds: cosmic ray (CR) events and hot pixels. Cosmic rays will generally not present a serious problem. The profile of M87 will be extremely well mapped by the observations so that all cosmic rays (except those that happen to look exactly like a PSF) will be easily recognized as such. That is, no CR splits will be necessary. While it is possible that a CR will occasionally mimic a PSF, it is extremely unlikely that this will happen in more than one consecutive observation. Hence the only real problem posed by CRs is that they will destroy the pixel signal in a small fraction of each image. Hot pixels could present a more serious problem. Although individual hot pixels are easily distinguished from a PSF, even on the highly undersampled WFC2, it is possible that a chance alignment of hot pixels could mimic a lensing event, since the response of hot pixels varies from exposure to exposure. This seems unlikely to me, but if tests show that the possibility is real, the telescope could be moved by one or two pixels between orbits.

Finally, there is the possibility that stellar events in M87 could mimic lensing. To do so, they would have to have a change in luminosity corresponding to $M_I \lesssim -6$ with a time symmetric light curve. I know of no stellar events in old metal-rich populations which would satisfy these criteria. Nevertheless, ongoing lensing searches toward the Galactic bulge and M31 should be studied to see if such events occur.

## 5. Other Parameters

In traditional microlensing searches (based on monitoring resolved stars) one recovers three parameters, $\omega$, $\beta$, and $t_0$ [see eqs. (2.1) and (2.3).] Only the time scale $\omega^{-1}$ provides information about the Macho itself, but this information is highly degenerate. An important problem is then to find additional information, either the proper motion $\mu$ (Gould 1994a; Nemiroff & Wickramasinghe 1994; Maoz & Gould 1994; Witt 1994) or the parallax (Gould 1992, 1994b, 1995b). However, even without these additional parameters one can determine the optical depth $\tau$ and the lensing rate $\Gamma_0$ directly from the data.

In pixel lensing one generally still measures three parameters, but the situation is less favorable. The measured parameters are the time of maximum magnification $t_0$, a characteristic time $\tilde{t}$, and the maximum extra luminosity, $\tilde{L}$, where

$$\tilde{t} = \frac{\beta}{\omega}, \qquad \tilde{L} = (A_{\max} - 1)L \simeq \frac{L}{\beta}. \tag{5.1}$$

Only $\tilde{t}$ gives information about the Macho, but this information is convolved with the impact parameter, itself a random variable. Hence one cannot directly recover the optical depth from pixel-lensing measurements. Instead, one measures only $\Gamma_0$.

It is important to obtain additional information for two reasons. First, as discussed in the previous section, one would like to distinguish the ICMs from the background generated by M87 stars. Second, one would like to constrain the properties of the ICMs themselves. Here I discuss two additional pieces of information: the color of the lensed star and the proper motion.

The color of the lensed star can be determined by making observations in two bands. Since the event itself is achromatic, equation (5.1) implies that the color can be determined from the ratio of two measurable quantities: $L_I/L_V = \tilde{L}_I/\tilde{L}_V$. The value of the color is that it allows one to determine approximately the luminosity of the star. The determination will be substantially better for brighter giants because magnitude is not such a steep function of color as it is for stars near the base of the giant branch. Nevertheless, even for the fainter stars the color does provide some information about the luminosity. To the extent that $L$ can be determined, $\beta$ can also be measured, and thus so can $\omega$ [see eq. (5.1)]. Hence, color measurements can be used to recover, or partly recover the optical depth. Because of scatter in the color-mag relation, colors need be accurate only to a few tenths. It should therefore not be necessary to make $V$ measurements more than a fraction of the time, say one in four or five measurements. Detailed simulation are needed to determine the trade-off between the accuracies of the basic $I$ band light curve and the color measurement.

If the lens transits the source, one can measure $q$, the ratio of the angle of projected closest approach of the Macho to the stellar radius (Gould 1994a; Nemiroff & Wickramasinghe 1994): $q \equiv \beta\theta_*/\theta_s$. Using this result and equation (5.1), one can therefore relate the proper motion $\mu$ to the stellar radius $\theta_s$ by means of measurable quantities:

$$\mu = \frac{q}{\tilde{t}}\theta_s. \qquad (5.2)$$

Since $\theta_s$ is a relatively weak function of the color, it should be possible to estimate $\mu$ fairly accurately. For essentially all the lenses, $D_{\rm OL} \simeq D_{\rm OS}$. Thus $\mu$ translates directly into a physical velocity: $v = \mu D_{\rm OS}$. Because the transverse speeds of ICMs are typically a factor $\sim 2$ faster than those of M87 stars, it should be possible to discriminate between the two populations, at least statistically. Recall that M87 stars must pass at least near to the limb of the source in order to give rise to a detectable event. This implies that in a substantial fraction of events the M87 stars will transit the source. A smaller, but non-negligible fraction of ICMs will also transit the source, particularly those detected in the inner $\sim 40''$ of M87.

## 6. Class II Events

In § 3, I estimated the number of events by requiring that a one-orbit measurement at peak yield a $6\,\sigma$ detection. In fact, there is a much larger class of events which are detectable, at least in principle, and which fail to satisfy this condition. To see this, consider again a source near the base of the giant branch ($M_I = 0.5$) at a point in M87 where the surface brightness is 19.5 mag arcsec$^{-2}$. Recall that for a typical threshold ($6\,\sigma$) event, $\beta = 1/160$ and $\tilde{t} \sim 2.4\,{\rm hr}$. Now consider another source that is five times brighter ($M_I \sim -1.25$). A $6\,\sigma$ detection can now be obtained at $\beta = 1/32$. Hence, in my earlier treatment, I assumed that the probability of a detectable lensing event is 5 times greater than for the $M_I = 0.5$ star. Now suppose that the impact parameter is five times higher yet, $\beta = 5/32$. The signal to noise achieved in a single orbit would only be $1.2\,\sigma$, far below my adopted threshold. However, the time scale of the event would be $25 \times 2.4\,\mathrm{hr} = 60\,\mathrm{hr}$. This long time scale implies that even if one binned the observations in groups of 25 orbits (and so increased the signal to noise to $6\,\sigma$) one would still resolve the temporal structure of the event as well as in the case of the $M_I = 0.5$ star with $\beta = 1/160$. Thus, if one demands only that there be a $6\,\sigma$ detection during a time $\tilde{t}$ close to the peak (instead of during a single orbit) then the number of detectable events is proportional to the square of a star's luminosity rather than to the luminosity itself. That is, equation (2.5) is replaced by

$$\Gamma = \Gamma_0 \frac{\Omega_{\mathrm{ccd}}}{\Omega_{\mathrm{psf}}} \eta \frac{\bar{L}}{L_-} \frac{t_-}{t_{\mathrm{orbit}}} \xi; \qquad \bar{L} \equiv \frac{\int dL\,\phi(L) L^2}{\int dL\,\phi(L) L}. \qquad (6.1)$$

Here $L_-$ is the luminosity of the fiducial star used previously (at the base of the giant branch $M_I = 0.5$), $\tilde{t}_-$ is the time scale for this star at the threshold value of $\beta$ (2.4 hr in the 19.5 mag arcsec$^{-2}$ field). The "fluctuation magnitude" $\bar{M}_I$ corresponding to the fluctuation luminosity $\bar{L}$, has been evaluated using the relation $\bar{M}_I = -4.84 + 3(V - I) = -1.18$ (Tonry 1991) where $(V - I)_{M87} = 1.22$ (J. Tonry private communication, 1995). The factor $\xi$ now includes a cutoff when $L^2 > \Delta L_{\min} L_- \tilde{t}_{\mathrm{orbit}}/t_-$. (There is also still a cutoff at the faint end, but this is usually unimportant.) If the factor $\xi$ is ignored then number of events in the "typical" field ($S = 19.5\,\mathrm{mag\,arcsec}^{-2}$) is enhanced by a factor $\sim 7.5$. Actually, the luminosity-function cutoff correspondends to $M_I = -2.25$, which is substantially brighter than the fluctuation magnitude and near the tip of the giant branch. Hence $\xi \sim 1$. In the field that is a factor 4 fainter, $\tilde{t}_-$ and is increased by a factor 2 while $\Delta L_{\min}$ is reduced by a factor 2. Hence the suppression factor becomes significant in the outer fields although it is non-existent in the inner fields. I therefore estimate that including these additional "Class II" events increases the overall event rate by a factor $\sim 6$.

From a purely statistical standpoint, Class II events are just as good as the Class I events described previously. However, there are two considerations that should make one cautious. First, the low signal to noise in the individual measurements (typically $\lesssim 1$) means that systematic errors can creep in rather easily. In an experiment which requires one to monitor millions of pixels for hundreds of observations, this can be an important drawback. Second, the Class II events will generally have impact parameters that are larger than the source stars. Hence, only a small fraction will have proper motions. This means that the main mechanism for discriminating between lensing events caused by ICMs and M87 stars will not work for the Class II events. Therefore, provided that these events can be reliably detected, they will be useful primarily to flesh out our understanding of the lens population, not to definitively detect ICMs.

## 7. Conclusion

It is feasible to detect Intra-Cluster Machos (ICMs) in the Virgo cluster by searching for pixel lensing events against M87. A dedicated one month search by *HST* should detect $\sim 40(f/20\%)$ Class I events where $f$ is the fraction of the Virgo cluster composed of Machos. An additional $\sim 30$ events would be detected due to lensing by stars in M87 itself. These could be distinguished from the ICMs by making proper motion measurements as the lens transits the face of the star. Class II events (where the total event is detected with high signal to noise, but none of the individual pixel magnifications rise above $6\sigma$) would occur 5 times more frequently than Class I events. For the most part, there would be no way to distinguish between Class II events due to ICMs and M87 stars.

Even if there are no ICMs, there should be a rate $\Gamma \sim 6\,\mathrm{day}^{-1}$ from M87 stars, including 1 Class I event and 5 Class II events. A three-day dedicated experiment could therefore be expected to detect $\sim 18$ events and would therefore prove or disprove the viability of the method.

**Acknowledgements**: I would like to thank P. Baillon, A. Crotts, J. Kaplan, and A. L. Melchior for stimulating discussions about their lensing searches toward M31.

# REFERENCES

Alcock, C., et al. 1993, Nature, 365, 621

Alcock, C., et al. 1995a, ApJ, 445, in press

Alcock, C., et al. 1995b, Phys. Rev. Lett., submitted

Aubourg, E., et al. 1993, Nature, 365, 623

Bahcall, J. N., Flynn, C., Gould, A. & Kirhakos, S. 1994, ApJ, 435, L51

Baillon, P., Bouquet, A., Giraud-Héraud, Y., & Kaplan, J. 1993, A&A, 277, 1

Crotts, A. P. S. 1992, ApJ, 399, L43

Faber, S. M. et al. 1989, APJS, 69, 763

Gould, A. 1992, ApJ, 392, 442

Gould, A. 1994a, ApJ, 421, L71

Gould, A. 1994b, ApJ, 421, L75

Gould, A. 1995a, ApJ, 441, 77

Gould, A. 1995b, ApJ Letters, in press

Gould, A. 1995c, ApJ Letters, submitted

Green, E. M., Demarque, P., & King, C. R. 1987, Revised Yale Isochrones and Luminosity Functions, (New Haven: Yale University Observatory)

Maoz, D. & Gould, A. 1994, ApJ, 425, L67

Nemiroff, R. J. & Wickramasinghe, W. A. D. T. ApJ, 424, L21

Tonry, J. L. 1991 ApJ, 373, L1

Udalski, A., Szymański, J., Stanek, K. Z. Kaluzny, J., Kubiak, M., Mateo, M. Krzemiński, W., Paczynski, B. & Venkat, R. 1994, Acta Astron, 44, 165

White, S. D. M. 1993, Nature, 366, 429

Witt, H. 1994, ApJ, submitted